\documentclass[num-refs]{nbdt-article}

\usepackage{siunitx}

\papertype{Original Article}
\paperfield{}

\title{Data Augmentation Through Monte Carlo Arithmetic Leads to More Generalizable Classification in Connectomics}

\author[1\authfn{2}]{Gregory Kiar PhD}
\author[2]{Yohan Chatelain PhD}
\author[2]{Ali Salari}
\author[1]{Alan C.~Evans PhD}
\author[2]{Tristan Glatard PhD}

\affil[1]{Montreal Neurological Institute, McGill University, Montreal, QC, H3A 2B4, Canada}
\affil[2]{Department of Computer Science and Computer Engineering, Concordia University, Montreal, QC, H3G 1M8, Canada}

\corraddress{Gregory Kiar PhD, Center for the Developing Brain, Child Mind Institute, New York,  NY, 10022, USA}
\corremail{gregory.kiar@childmind.org}

\presentadd[\authfn{2}]{Center for the Developing Brain, Child Mind Institute, New York,  NY, 10022, USA}

\fundinginfo{Natural Sciences and Engineering Research Council of Canada (NSERC) 
, Award Number: CGSD3-519497-2018; In partnership with the Canadian Open Neuroscience Platform.}

\runningauthor{G. Kiar et al.}

\begin{document}
\emergencystretch 3em

\maketitle

\begin{abstract}
Machine learning models are commonly applied to human brain imaging datasets in an effort to associate function or structure with behaviour, health, or other individual phenotypes. Such models often rely on low-dimensional maps generated by complex processing pipelines. However, the numerical instabilities inherent to pipelines limit the fidelity of these maps and introduce computational bias. Monte Carlo Arithmetic, a technique for introducing controlled amounts of numerical noise, was used to perturb a structural connectome estimation pipeline, ultimately producing a range of plausible networks for each sample. The variability in the perturbed networks was captured in an augmented dataset, which was then used for an age classification task. We found that resampling brain networks across a series of such numerically perturbed outcomes led to improved performance in all tested classifiers, preprocessing strategies, and dimensionality reduction techniques. Importantly, we find that this benefit does not hinge on a large number of perturbations, suggesting that even minimally perturbing a dataset adds meaningful variance which can be captured in the subsequently designed models.

% Please include a maximum of seven keywords
\end{abstract}

\section{Introduction}
The application of machine learning has become commonplace for the identification and characterization of individual biomarkers in neuroimaging~\cite{woo2017building}. Models have been applied to discriminate between measures of brain structure or function based upon phenotypic variables related to disease~\cite{Crossley2014-tg,Payabvash2019-tm,Tolan2018-nq}, development~\cite{Zhang2019-ko}, or other axes of potential consequence~\cite{Zhu2018-cs,Park2015-uj}.

These models often build representations upon processed imaging data, in which 3D or 4D images have been transformed into estimates of structure~\cite{wade2017machine}, function~\cite{weis2020sex}, or connectivity~\cite{munsell2015evaluation}. However, there is a lack of reliability in these estimates, including variation across analysis team~\cite{botvinik2020variability}, software library~\cite{bowring2019exploring}, operating system~\cite{salari2020file}, and instability in the face of numerical noise~\cite{Kiar2020-kz}. This uncertainty limits the ability of models to learn generalizable relationships among data, and leads to biased predictors. Traditionally, this bias has been reduced through the collection and application of repeated-measurement datasets~\cite{sudlow2015uk,zuo2014open}, though this requires considerable resources and is not feasible in the context of all clinical populations.

Dataset augmentation has been broadly demonstrated as an effective technique for improving the generalizability of learned models, especially in deep learning~\cite{nalepa2019data}. In neuroimaging, augmentation is often performed throughout the process of tool validation and requires either realistic data simulators or phantom datasets~\cite{graham2016realistic}, which significantly limits their applicability to real data. Recent advances in deep learning have made dataset augmentation far more accessible~\cite{shorten2019survey}, including in medical imaging~\cite{yi2019generative,barile2021data}. However, due to the lack of widely accepted noise models in connectomics, the necessity of compute-intensive training, and the associated infrastructural requirements, the application of dataset augmentation to network neuroscience remains largely uncharted territory (see Appendix~1\ref{app:augmentation} for details).

Perturbation methods which inject small amounts of noise through the execution of a pipeline, such as Monte Carlo Arithmetic (MCA)~\cite{Parker1997-qq,Denis2016-wo}, have recently been used to induce instabilities in structural connectome estimation software~\cite{Kiar2020-lb}. Importantly, this technique produces a range of equally plausible results, where no single observation is more or less valid than the others – including those which were left unperturbed. While sampling from a set of perturbed connectomes may have an impact on learning brain-phenotype relationships~\cite{Kiar2020-kz}, there remains potential for leveraging the distribution of perturbed results to augment datasets in lieu of increasing sample sizes, performing repeated measurements, or developing data-specific augmentation models.

Using an existing collection of MCA-perturbed structural connectomes~\cite{Kiar2020-yz}, we trained classifiers on networks sampled from the distribution of results and evaluated their performance relative to using only the unperturbed networks. We evaluate several techniques for resampling the networks, and compare classifiers through their validation performance, the performance on an out-of-sample test set, and the generalizability of their performance across the two. We demonstrate the efficacy of using MCA as a method for dataset augmentation which leads to more robust and generalizable models of brain-phenotype relationships.

\section{Materials \& Methods}
\begin{figure}[bth!]\centering
\includegraphics[width=\linewidth]{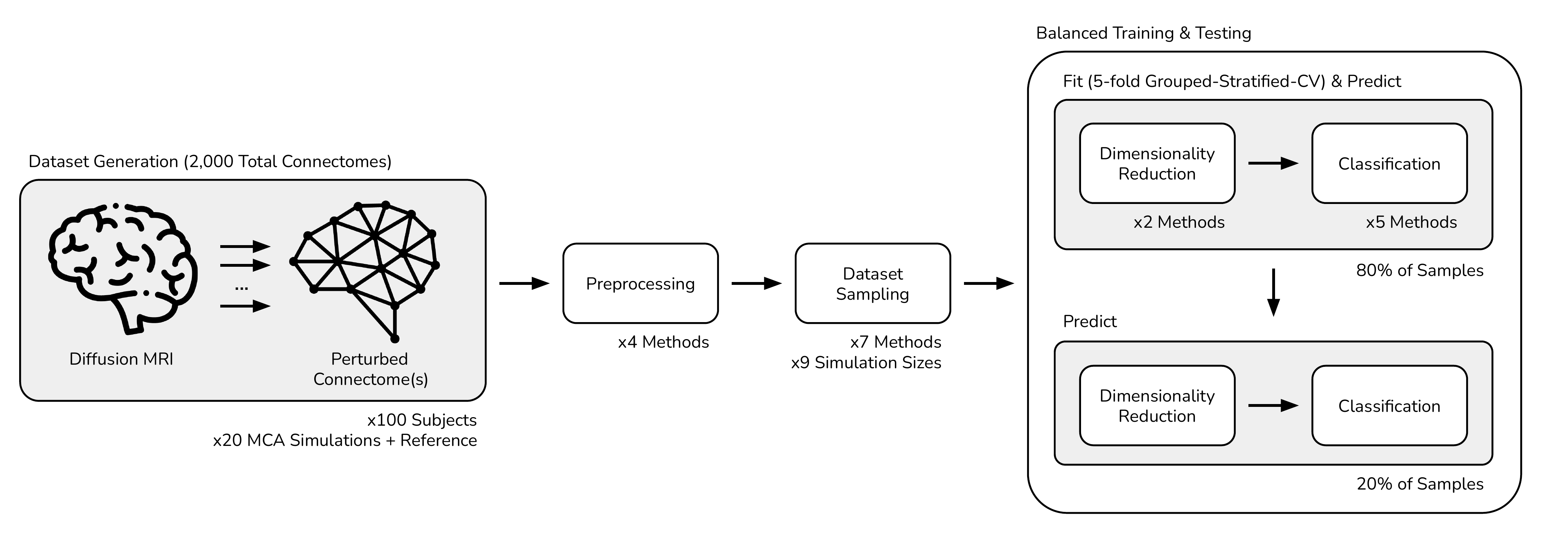}
\caption{Experiment workflow. MCA-simulated connectomes were previously generated for $100$ subjects, $20$ times each. The resulting dataset of $2,000$ connectomes were independently preprocessed using one of $4$ techniques. The dataset was then sampled according to one of $7$ dataset sampling strategies and using one of $9$ possible number of MCA simulations per subject. The dataset was split into balanced training and testing sets, and fed into the models. The models consisted of one of $2$ dimensionality reduction techniques prior to classification using one of $5$ classifier types. The models were fit and made predictions on the training set prior to making predictions on the test set.}
\label{fig:workflow}
\end{figure}

The objective of this study was to evaluate the impact of aggregating collections of unstable brain networks towards learning robust brain-phenotype relationships. We sampled and aggregated simulated networks within individuals to learn relationships between brain connectivity and individual phenotypes, in this case age, and compared this to baseline performance on this task. We compared aggregation strategies with respect to baseline validation performance, performance out-of-sample, and generalizability. The experimental workflow is visualized in Figure~\ref{fig:workflow}.

\noindent All developed software and analysis resources for this project have been made available through GitHub at:

\url{https://github.com/gkpapers/2021AggregateMCA}.

\subsection{Dataset}
An existing dataset containing Monte Carlo Arithmetic (MCA) perturbed structural human brain networks was used for these experiments~\cite{Kiar2020-yz}. While further information on the processing and curation of this dataset can be found in~\cite{Kiar2020-kz}, a brief description of the data and processing follows here.

The networks were derived from diffusion MRI data from the Nathan Kline Institute Rockland Sample
dataset~\cite{zuo2014open}. The data were denoised and aligned prior to undergoing modelling via a probabilistic tractography pipeline built using Dipy~\cite{Garyfallidis2014-ql} with a fixed random state. The streamlines were ultimately mapped to regional network connections using the Desikan-Killiany-Tourville parcellation~\cite{Klein2012-vi}. The raw input data to this pipeline consisted of $4$-dimensional images (with the fourth dimension corresponding to unique diffusion directions) containing O($10^8$) voxels, and the derived connectomes were matrices with dimensions $83 \times 83$, where each value at location ($i$, $j$) corresponds to the strength of connection between region $i$ and region $j$. As connections are undirected (i.e. the edge ($i$, $j$) is equal to the edge ($j$, $i$)), the matrices each consist of $3,403$ unique weighted connections.

Perturbations were introduced for the diffusion modelling of this dataset using sparsely-introduced Monte Carlo Arithmetic at the level of machine-precision, simulating expected error over a typical pipeline execution. The MCA workflow was such that each inexact floating-point operation was perturbed with a zero-centered random variable 1-bit beyond the precision of the system~\cite{Denis2016-wo,Parker1997-qq}. In practice, this resulted in perturbations that were analogous to an inexact trailing decimal value of, say, $0.6$ being rounded up to $1.0$ in $60\%$ of evaluations and down to $0.0$ for the remaining $40\%$. The perturbations were randomly sampled for each operation, and thus the unique combinations of error accumulation over multiple executions of a pipeline lead to potentially distinct terminal results. The MCA approach is pipeline- and data-agnostic, making it widely applicable to a variety of pipelines or domains which may suffer a lack of context-specific perturbation or dataset augmentation strategies.

This dataset contains a single session of data from $100$ individuals ($100\times 1 \times1$). The number of subjects included is consistent for the types of analyses performed below, and considered large in general for neuroimaging~\cite{neuroimagingsamplesize}. Each sample was simulated $20$ times, resulting in $2,000$ unique graphs. This collection enabled the exploration of subsampling and aggregation methods in a typical learning context for neuroimaging~\cite{Dimitriadis2017-pd,Buchanan2014-pm}. Exploring the relationship between the number of simulations and performance further allows for the cost of MCA-enabled resampling to be evaluated in the context of dataset augmentation.

As the target for classification, individual-level phenotypic data strongly implicated in brain connectivity was desired. Participant age, which has consistently been shown to have a considerable impact on brain connectivity~\cite{Meier2012-ve,Wu2012-uc,Bookheimer2019-ti,Zhao2015-rm}, was selected and turned into a binary target by dividing participants into adult ($>18$) and non-adult groups ($68\%$ adult). For subsequent validation of findings, body mass index was chosen as a second target ($50\%$ overweight).

\subsection{Preprocessing}
Prior to being used for this task, the brain networks were represented as symmetric $83 \times 83$ adjacency matrices, sampled upon the Desikan-Killiany-Tourville~\cite{Klein2012-vi} anatomical parcellation. To reduce redundancy in the data, all edges belonging to the upper-triangle of these matrices were preserved and vectorized, resulting in a feature vector of $3,403$ edges per sample. All samples were preprocessed using one of four standard techniques:

\paragraph{Raw} The raw streamline count edge-weight intensities were used as originally calculated.

\paragraph{Log Transform} The log10 of edge weights was taken, and edges with $0$ weight prior to the transform were reset to $0$.

\paragraph{Rank Transform} The edges were ranked based on their intensity, with the largest edge having the maximum value. Ties were settled by averaging the rank, and all ranks were finally min-max scaled between $0$ and $1$.

\paragraph{Z-Score} The edge weights were z-scored to have a mean intensity of $0$ and unit variance.

\subsection{Machine Learning Pipelines}

The preprocessed connectomes were fed into pipelines consisting of two steps: dimensionality reduction and classification. Given the high dimensionality ($3,403$ features) of the networks, the relatively small number of samples, and standard practice in the field~\cite{Payabvash2019-tm,Park2015-uj,weis2020sex}, no experiments were considered without dimensionality reduction. Dimensionality reduction was applied using one of two methods:

\paragraph{Principal Component Analysis} The connectomes were projected into the $20$ dimensions of highest variance. The number of components was chosen to capture approximately $90\%$ of the variance present within the dataset.

\paragraph{Feature Agglomeration} The number of features in each connectome were reduced by combining edges according to maximum similarity/minimum variance using agglomerative clustering~\cite{Ward1963-uh}. The number of resulting features was $20$, to be consistent with the number of dimensions present after PCA, above.

After dimensionality reduction, samples were fed into one of five distinct classifiers as implemented through scikit learn~\cite{Pedregosa2011-uz}:

\paragraph{Support Vector Machine} The model was fit using a radial basis function (RBF) kernel, L2 penalty, and a balanced regularization parameter to account for uneven class membership.

\paragraph{Logistic Regression} A linear solver was used due to the relatively small dataset size. L2 regularization and balanced class weights were used, as above.

\paragraph{K-Nearest Neighbour} Class membership was determined using an L2 distance and the nearest $10\%$ of samples, scaling with the number of samples used for training.

\paragraph{Random Forest} $100$ decision trees were fit using balanced class weights, each splitting the dataset according to a maximum of $4$ features per node (corresponding to the rounded square root of $20$ total features).

\paragraph{AdaBoost} A maximum of $50$ decision trees were fit sequentially such that sample weights were iteratively adjusted to prioritize performance on previously incorrectly-classified samples, consistent with~\cite{Freund1997-qy}.

The hyperparameters for all models were refined from their default values to be appropriate for a small and imbalanced dataset. The performance for all pipeline combinations of preprocessing methods, dimensionality reduction techniques, and models using the reference (i.e. unperturbed) executions in the dataset ranged from an F1 score of $0.64 - 0.875$ with a mean of $0.806$; this evaluation was performed on a consistent held-out test set which was used for all experiments, as described in a following section. This set of models was chosen as it includes i) well understood standard techniques, ii) both parametric and non-parametric methods, iii) both ensemble and non-ensemble methods, and iv) models which have been commonly deployed for the classification of neuroimaging datasets~\cite{Meier2012-ve,Tunc2016-cz,Zhu2018-cs,Payabvash2019-tm,Crossley2014-tg,Park2015-uj,Nayak2016-wl,Tolan2018-nq}.

\subsection{Dataset Sampling}

A chief purpose of this manuscript involves the comparison of various forms of aggregation across equivalently-simulated pipeline outputs. Accordingly, the dataset was resampled after preprocessing but prior to dimensionality reduction and classifiers were trained, evaluated, and combined according to the following procedures:

\paragraph{Reference} Networks generated without any MCA perturbations were selected for input to the models, serving as a benchmark.

\paragraph{Truncate} The number of significant digits~\cite{Parker1997-qq} per-edge was calculated using all simulated networks, and the edge weights in the reference graph were truncated to the specified number of digits. Importantly, this is the only method used which deliberately squashes the variance observed across simulations.

\paragraph{Jackknife} The datasets were repeatedly sampled such that a single randomly chosen observation of each unique network was selected (i.e. derived from the same input datum). This resampling was performed $100$ times, resulting in the total number of resamplings being $5\times$ larger than the number of unique observations per network, ensuring a broad and overlapping sampling of the datasets.

\paragraph{Median} The edgewise median of all observations of the same network were used as the samples for training and evaluation.

\paragraph{Mean} Similar to the above, the edgewise mean of all observations for each network were computed and used as input data to the classifiers in both collections.

\paragraph{Consensus} A distance-dependent average network~\cite{Betzel2018-eo} was computed across all observations of each network. This data-aware aggregation method, developed for structural brain network analysis, preserves network properties often distorted when computing mean or median networks.

\paragraph{Mega-analysis} All observations of each network were used simultaneously for classification, increasing the effective sample size. Samples were organized such that all observations of the same network only appeared within a single fold for training and evaluation, ensuring double-dipping was avoided.

\paragraph{Meta-analysis} Individual classifiers trained across jackknife dataset resamplings, above, were treated as independent models and aggregated into an ensemble classifier. The ensemble was fit using a logistic regression classifier across the outputs of the jackknifed classifiers to learn a relationship between the predicted and true class labels.

The robustness and possible benefit of each subsampling approach was measured by evaluation on a subset of all MCA simulations, including $9$ distinct numbers of simulations, ranging from $2$ to $20$ simulations per sample. Combining the dataset sampling methods, the set of simulations, preprocessing strategies, dimensionality reduction techniques, and classifier models, there were $2,520$ perturbed models trained and evaluated next to $40$ reference models.

\subsection{Training \& Evaluation}

Prior to training models on the brain networks, $20\%$ of subjects were excluded from each dataset for use as an out-of-sample test dataset for all experiments. With the remaining $80\%$ of subjects, cross validation was performed following a stratified grouped $k$-fold approach ($k=5$). In this approach, samples were divided into training and validation sets such that the target variable was proportionally represented on each side of the fold (stratified), conditional upon all observations from the same individual, relevant for the mega-analysis dataset sampling method, falling upon the same side of the fold (grouped). This resulted in $5$ fold-trained classifiers per configuration, each trained on $64\%$ of the samples and validated on $16\%$, prior to each being tested on the remaining $20\%$ of held-out samples. All random processes used in determining the splits used the same seed to remove the effect of random sampling.

Classifiers were primarily evaluated on both the validation and test (out-of-sample) sets using F1 score, a standard measure for evaluating classification performance. The generalizability of predictions was defined as:

\begin{equation}
G = 1 - \lvert F1_{test} - F1_{validation} \rvert
\label{eq:gen}
\end{equation}

where a score of $1$ (maximum) indicates the equivalent performance across both the validation and test sets, and a lower score (minimum of $0$) indicates inconsistent performance. The absolute change in performance was used in Eq.~\ref{eq:gen}, resulting in a score which penalizes spurious over-performance similarly to under-performance. This is a desired attribute of the measure as all inconsistency, whether due to chance or model fit, is undesirable when applying a classifier out-of-sample. Importantly, this measure does not use training performance as the reference, as is common in deep learning, but instead considers the validation performance. This is because training performance is prone to unrealistic inflation due to dataset memorization which may arise in some classifiers; K-Nearest Neighbours is a good example of this. Given that the validation performance is typically reported on classic models, this can be considered a \textit{practical} definition of generalizability, and is still in the spirit of ``evaluating performance on seen data to that on unseen data''~\cite{shorten2019survey}.

Differences in F1 score and generalizability for perturbed experiments with respect to their reference were used to measure the change in performance for each dataset sampling technique, and statistical comparisons were made through Wilcoxon Signed-Rank tests.

\section{Results}

\begin{table}[b!]
\centering
\caption{Statistically significant change in performance. Red values indicate significant
decline in performance, black values indicate improvement, and empty cells indicate no change. A single star
represents $p < 0.05$, and each additional star is an additional order of magnitude of significance.}
\label{tab1:perf}
\small
\begin{tabular}{rccc}
\textbf{Dataset Sampling}  & \textbf{Validation} &      \textbf{Test} & \textbf{Generalizability} \\
\hline
\textbf{Truncate}          &    {\color{red} **} &                    &                           \\
\textbf{Jackknife}         &    {\color{red} **} &                 ** &                           \\
\textbf{Mean}              &                     &                *** &                           \\
\textbf{Median}            &                     &                *** &                           \\
\textbf{Consensus}         &                     &                *** &                         * \\
\textbf{Mega-Analysis}     &     {\color{red} *} &                  * &                       *** \\
\textbf{Meta-Analysis}     &                  ** &                *** &                         * \\
\end{tabular}

\end{table}

The figures and findings presented in this section represent a summary of the complete experiment table which consists of performance measures and metadata for all $2,560$ models tested. The complete performance table alongside the table of significant differences, are made available through the GitHub repository.

\subsection{Data Resampling Improves Classification}

The overall performance of each subsampling method is summarized in Table~\ref{tab1:perf}. The change in performance was measured in both cases as a change in F1 score on the validation set, the change in F1 score on the test set, and the change in overall generalizability, a measure which summarizes the similarity between validation and test performance for a given model (Eq.~\ref{eq:gen}).

\begin{figure}[bth!]\centering
\includegraphics[width=\linewidth]{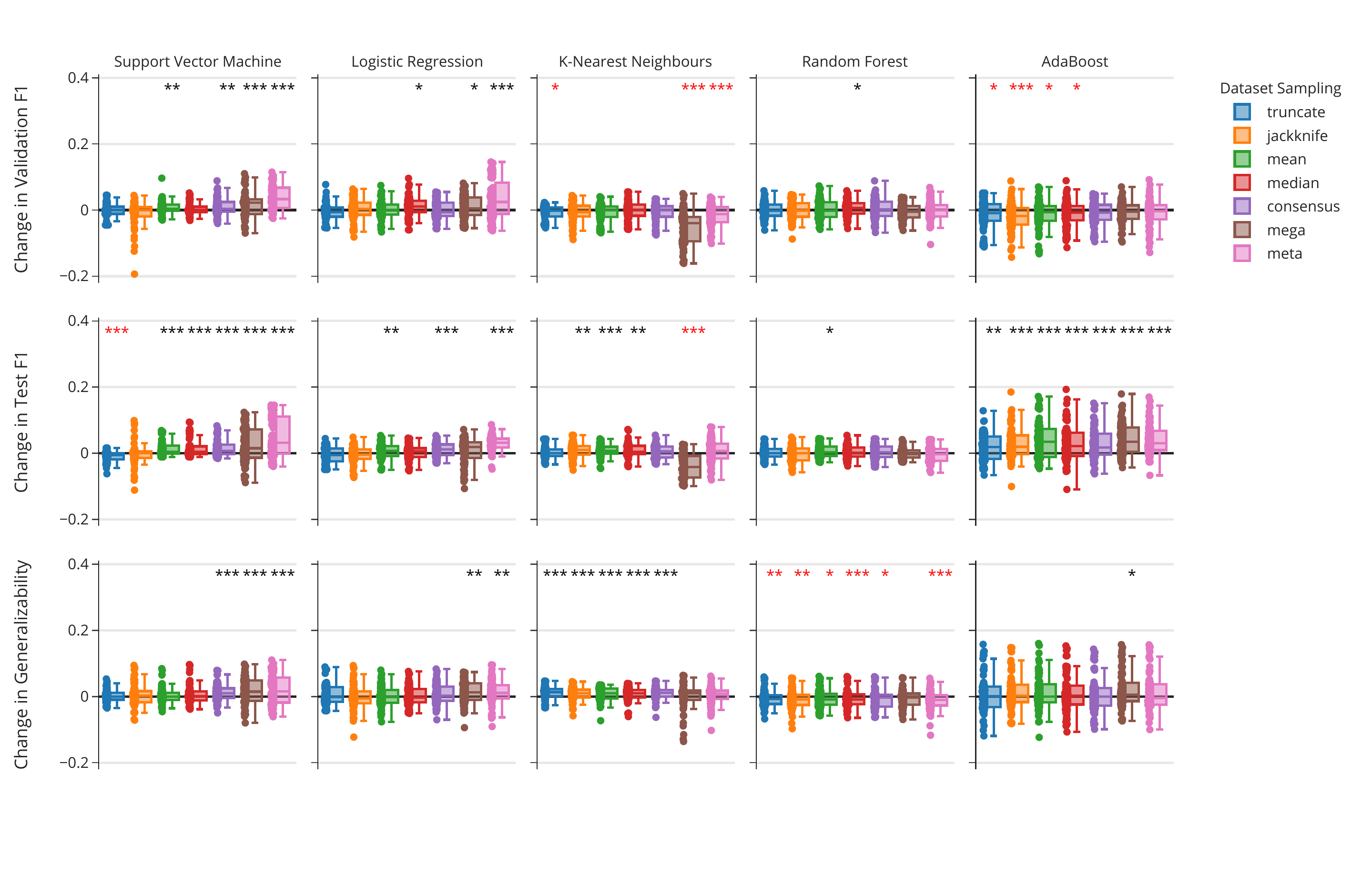}
\caption{Relative change in classifier performance with respect to classifier type and dataset sampling strategies as measured by change in F1 score on the validation set (top) or test set (middle), as well as the generalizability of performance (bottom). Each star annotation indicates an order of magnitude of statistically significant change, beginning at $0.05$ for one star and decreasing from there, with those in black or red indicating an increase or
decrease due to resampling, respectively.}
\label{fig:overall_perf}
\end{figure}

Across all classifier types it was found that consensus, mega-, and meta-analytic approaches outperformed other dataset resampling methods. The dataset sampling methods led to improved testing performance and generalizability on average from $0.773$ to $0.812$ and from $0.960$ to $0.965$, respectively. These results were similar to those found in the prediction of body mass index, which noted $0.033$ and $0.017$ improvements in F1 and generalizability on average, respectively. Further detail on the results associated with body mass index prediction can be found in Appendix~2. The noted improvement for both tasks is consistent with other dataset augmentation approaches mentioned in the following section, which typically report improvements from $0.01-0.1$~\cite{shorten2019survey}. In some cases, such as several configurations of AdaBoost classifiers, improvements of up to $0.170$ in F1 and $0.150$ in generalizability were simultaneously observed, with model generalizability peaking at $0.994$.

The only method which did not improve performance at all was the truncation resampling approach. This method was distinct from the others in that the variance observed across simulations was used to estimate and squash variance in the reference network, whether other approaches captured the variance. The finding that truncation hurts performance importantly suggests that the variability across the observed networks contains biologically meaningful signal.

The change in performance for each model and dataset sampling technique is shown in Figure~\ref{fig:overall_perf}. The support vector machine and logistic regression models improve across each of the three measures for a variety of dataset sampling techniques, suggesting that the addition of the MCA-perturbed samples improves the training, testing, and overall generalizability of these classifiers.

Distinctly, k-nearest neighbours (KNN) and AdaBoost classifiers experienced minimal change in validation and often saw their performance decline. However, the improvement of these classifiers on the test set suggests that resampling reduced overfitting in these classifiers. In the case of KNN, this translates to improved generalizability, while in the case of AdaBoost generalizability was largely unchanged, suggesting that the model went from underperforming to overperforming after dataset resampling. The unique decline in performance when using the mega-analytic resampling technique on KNN classifier is suggestive of poor hyper-parameterization, as there is a strong relationship between the performance and the ratio of the number of samples in the dataset to the $k$ parameter of the model. At present this parameter was scaled linearly with the number of MCA simulations used, however, it is both possible that an improved scaling function exists or that the model performance degrades with large sample sizes making it a poor model choice given this resampling technique.

The random forest classifiers uniquely did not see a significant change in validation or testing performance across the majority of resampling techniques. However, these classifiers did experience a significant decrease in the generalizability of their performance, meaning that there was a larger discrepancy between training and testing performance in many cases. This distinction from the other models is possibly due to the fact that random forest is a simple ensemble technique which takes advantage of training many independent classifiers and samples them to assign final class predictions, allowing this approach to form more generalizable predictions, and thus the addition of more data does not significantly improve performance further. While AdaBoost is also an ensemble method, the iterative training of models with increasing emphasis on difficult samples allows for the added variance in those samples to play an increasingly central role in the construction of class relationships, and thus more directly takes advantage of the added variability in samples.

While certain combinations of preprocessing, dimensionality reduction, and classifiers performed more harmoniously than others, there was no significant relationship between the performance of any single resampling method and preprocessing or dimensionality reduction technique. Overall, the above results show that dataset augmentation through MCA-perturbed pipeline outputs may be an effective way to improve the performance and generalizability of non-ensemble classifiers tasked with modelling brain-phenotype relationships, both within and out of sample, especially when variance is captured rather than removed.

\begin{figure}[t!]\centering
\includegraphics[width=\linewidth]{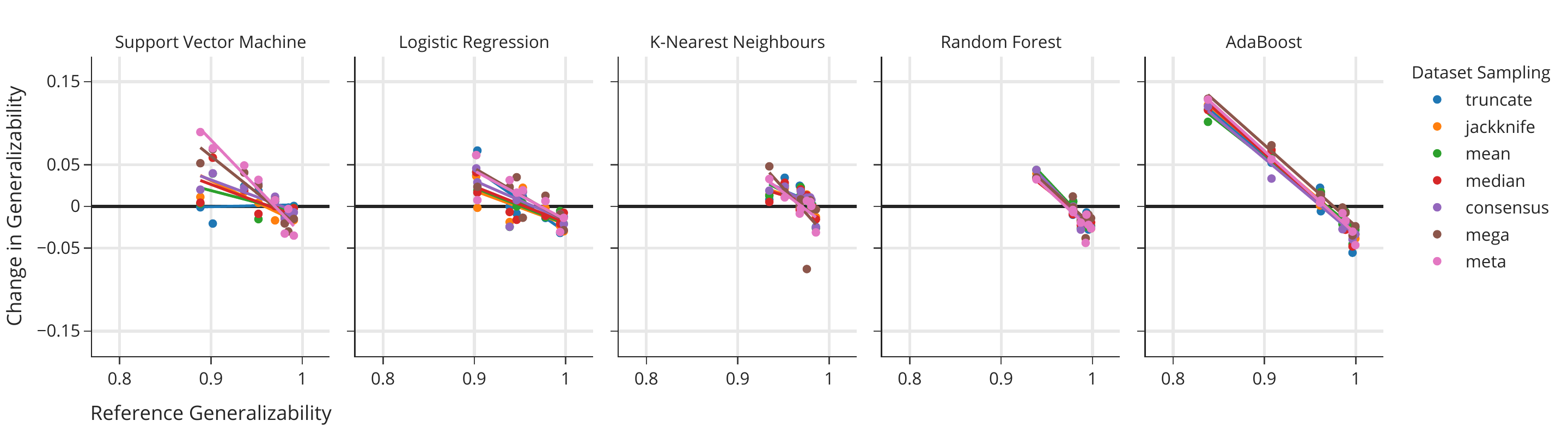}
\includegraphics[width=\linewidth]{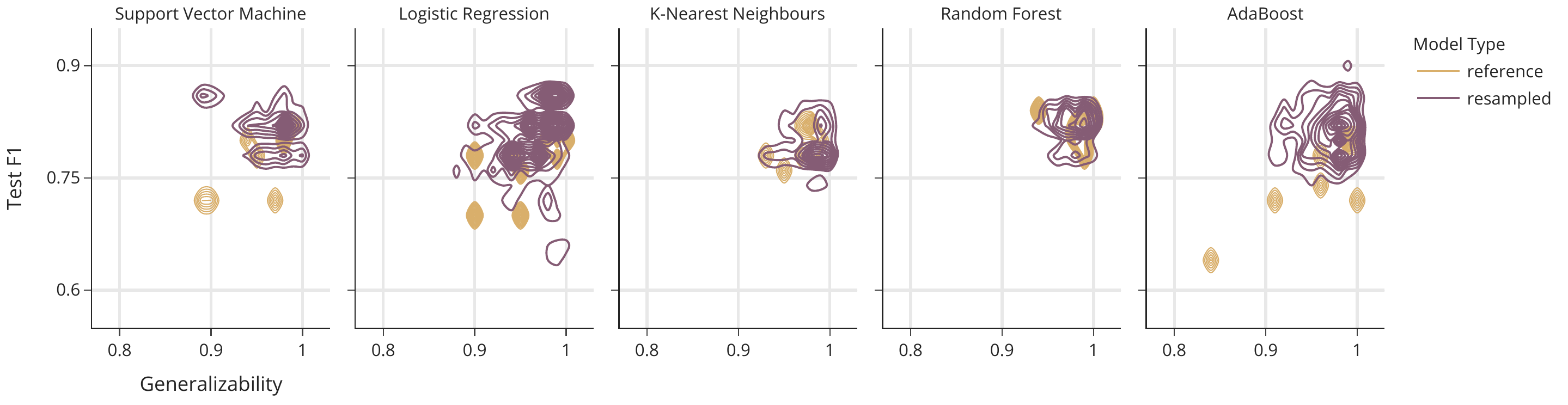}
\caption{Relationship between generalizability and resampling. Top: change in the generalizability of classifiers with respect to the reference generalizability. Each data point represents the mean change in generalizability for all models using the same preprocessing and dimensionality reduction techniques for a given classifier and dataset sampling strategy. Bottom: contour density distributions of generalizability and F1 scores across all models for both reference and resampled training.
% The vertical lines for each plot represent the mean generalizability for the associated models.
}
\label{fig:change_in_gen}
\end{figure}

\subsection{Resampling Leads to Consistent Performance}
To better characterize the benefit of resampling, the relationship between the magnitude of improvement and the baseline performance of the classifier were further explored (Figure~\ref{fig:change_in_gen}). We found that the increase in the generalizability of a classifier was inversely related to the baseline generalizability (Figure~\ref{fig:change_in_gen}; top). In other words, the less generalizable a classifier was originally, the more its generalizability improved (significant at $p < 0.05$ for all dataset sampling strategies and classifier other than KNN). There were several situations in which the generalizability of models were noted to decrease, however, though this only occurred for models with high generalizability scores (all $>0.935$). Importantly, the relative change in generalizability shifts scores towards a single ``mode'', suggesting a less biased estimate of the true generalizability of performance on the task, and mitigating both under- and over-performance due to chance.

When exploring the relationship between F1 and generalizability (Figure~\ref{fig:change_in_gen}; bottom), it becomes apparent that even for the majority of models which may not have improved performance along both axes, either their generalizability or F1 score is improved. While an ideal classifier would reside in the top-right of the shown plots, the dataset resampling techniques consistently shift the distributions in this direction and often improve classifiers along one or both of these axes. Importantly, the variance in performance across both measures is significantly decreased, suggesting that resampling leads to more reliable and reproducible classifiers.

\begin{figure}[t!]\centering
\includegraphics[width=\linewidth]{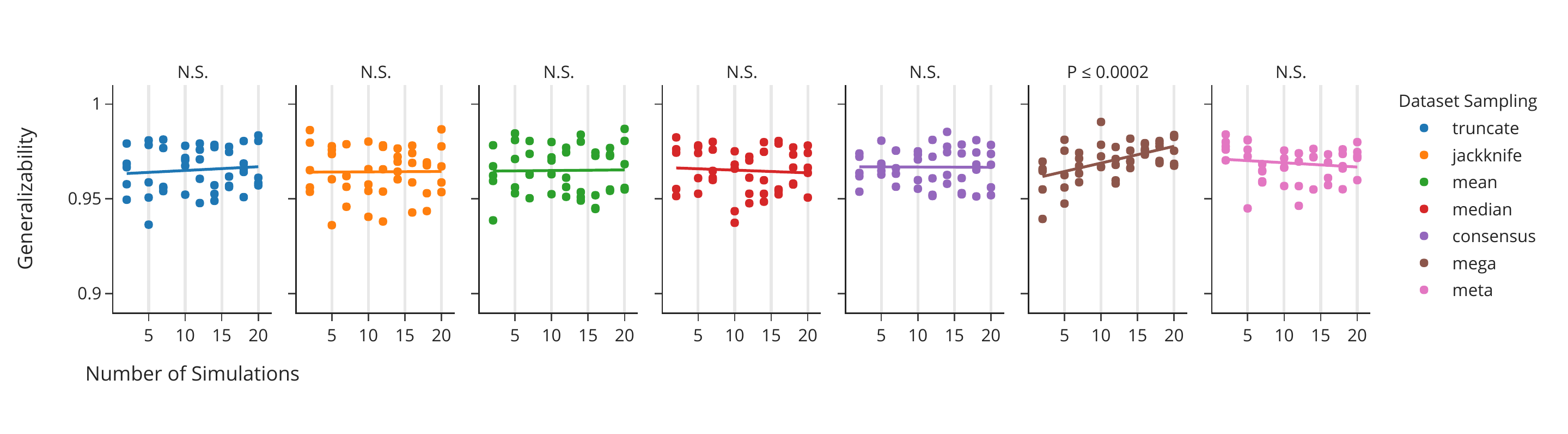}
\caption{The generalizability of classifiers using each dataset sampling technique with respect to the number of MCA simulations. Each number of simulations was sampled a single time, to avoid artificial skewing of the dataset due to the inclusion of ``higher'' or ``lower'' quality samples; a single drawing of each split mimics a true perturbation experiment context.}
\label{fig:nmca}
\end{figure}

\subsection{Number of Simulations is Largely Unimpactful}
While we previously noted an increase in classifier performance due to perturbation-enabled dataset resampling, it was important to explore the relationship between the number of simulated samples and performance (Figure~\ref{fig:nmca}). There was no relationship between the number of independent simulations and performance, as measured by either F1 or generalizability, for all dataset resampling techniques other than mega-analysis. In the case of the mega-analytic approach, however, there was a significant positive relationship between the number of samples used and the generalizability of performance, though there remained no increase in F1 score. The mega-analysis approach is the only approach which changes the number of samples being provided directly to the classifiers, thus mimics an increase in sample size for traditional experiments. While outlying samples may play a small role in many of the proposed forms of resampling, or non-existent in the median case, the mega analytic approach treats all simulations with equal importance as unique samples in the dataset. In this case, the relationship we observe is consistent to what one might expect when increasing the number of samples.

\section{Discussion}

The numerical perturbation of analytic pipelines provides a unique, data-agnostic, and computationally unintrusive method for dataset augmentation. Using a technique such as MCA, samples can be simulated across an array of controlled executions and used to enrich datasets with a range of plausible results per sample. We demonstrate that this method of dataset augmentation can be used to improve the training, testing, and generalizability of classifiers.

Through the training and evaluation of $2,560$ models combining varying strategies for preprocessing, dimensionality reduction, classifier, and resampling, we found consistent improvement across all measured axes. Interestingly, while there was a statistically significant improvement when using many dataset resampling techniques, there was no significant improvement in the performance, and in fact a reduction, using the truncation resampling method as is shown in Table~\ref{tab1:perf}. This result importantly demonstrates that the added variability in results obtained through MCA is valuable and signal-rich itself, and an important determination of performance is the inclusion of this variability. Recent work has demonstrated another impactful source of variability in machine learning benchmark performance, namely the initialization and parameterization of models~\cite{bouthillier2021accounting}. Our studies reached a common conclusion, which is that model performance can be improved by considering a variety of individually-biased estimators. These results are both elegant realizations of the bias-variance trade-off~\cite{geman1992neural}.

While the non-ensemble methods benefited most obviously from the dataset resampling strategies, where both F1 and generalizability were often improved, the results presented in Figure~\ref{fig:change_in_gen} demonstrate that variability in performance across both of these axes was reduced across all classifier configurations. While a reduction in the variability of performance is desirable in itself, this figure also illustrates that the performance of resulting models converges around the more performant models in the case of all classifiers. The reduction in variability also results in models which differed less significantly when looking across classifier types.

Although performance was improved by the integration of MCA simulated samples, performance was not significantly related to the number of simulations used in any case other than the mega-analytic resampling strategy. The independence of performance and number of simulations is encouraging, as a key limitation for using Monte Carlo methods is the often extreme computational overhead. The ability to use a small number of simulations and achieve equivalent performance through the majority of resampling techniques allows for significantly better performance without added data collection and only a theoretical doubling the sample processing time. The benefit of increasing the number of simulations in the mega-analytic case could be considered an analog to increasing the sample size of an experiment. While the range of simulations used here demonstrated a consistent improvement in generalizability, there will be a plateau in performance, either at a perfect score or, more likely, before this is reached. Further work is required for characterizing the relationship between the performance of mega-analytic resampling and the number of simulations, though it is likely that this relationship will be domain-specific and dependent on other experimental design variables such as the number of features per sample.

While our study shows that classifiers with poorer baseline performance benefit more from augmentation, an important limitation of this is the operating point to which that claim remains true. For example, it is unlikely that the trend observed here for a task with a mean reference performance of $0.81$ would hold across models operating with reference performance near chance or near perfect. Characterizing the behaviour of this technique across a range of classification contexts and performances would shed light on whether this technique could be applied globally or if it is limited to making ``good'' models better.

It is a well understood problem that small sample sizes lead to uncertainty in modelling~\cite{varoquaux2018cross}. This is generally planned for in one of two ways: the collection of vast datasets, as is the case in the UK-BioBank which strives to collect samples from half a million individuals~\cite{sudlow2015uk}, or the collection of repeated measurements from the selected samples, as is the case in the Consortium of Reliability and Reproducibility which orchestrates multiple centres and scanners, each collecting multiple acquisitions~\cite{zuo2014open}. In either case, the additional data collection by these initiatives is both financially and temporally expensive and leads to unintended confounding effects associated with time of day~\cite{vandewalle2009functional}, weather~\cite{di2019estimations}, or other off-target variables that may be poorly described in the resulting dataset~\cite{chaddock2010neuroimaging}.

While the results presented here provide strong evidence in favour of dataset augmentation through numerical perturbations, the improvement from these methods has not been directly compared to additional data acquisitions in this experiment due to the limited sample size of the available perturbed dataset~\cite{Kiar2020-yz}. Previous studies exploring the effect of sample size on neuroimaging classification tasks have shown that variability in performance decreases with sample size~\cite{chu2012does}, where a doubling of sample size from $100$ to $200$ approximately corresponded to halving the uncertainty in performance~\cite{varoquaux2018cross}. However, this decrease in variability is often accompanied by a decrease in out of sample performance in practice~\cite{schnack2016detecting}. A meta-analysis across $69$ studies showed that increasing sample size was negatively related to out-of-sample performance~\cite{pulini2019classification}, where accuracy was noted to decline by approximately $5\%$ in a similar doubling from $100$ to $200$ samples, suggesting that a major contribution of increasing sample size in neuroimaging is a reduction in overfitting which must occur prior to a possible boost in performance. Our finding that MCA-based dataset augmentation reduced overfitting and improved upon baseline performance is encouraging, and suggests that models trained using such perturbed datasets may benefit more from increased data collection.

A common issue in many machine learning contexts is the unbalanced nature of datasets. When using a nearest-neighbour classifier, for instance, a dramatic difference in the membership of each group could have significant impact on model hyper-parameters and performance. In contexts where balanced sampling is not possible, such as when considering a rare clinical population, perturbation-augmented datasets could be applied for realistic upsampling of data. In this case, a mega-analytic aggregation strategy could be used in which more simulations would be performed for members of the under-represented class, similar to the balancing of weights applicable to some models. This application is particularly important, as upsampling is often challenging in biological contexts where realistic simulation models are sparse.

The presented dataset augmentation technique was not compared directly to other strategies due to a lack of readily available alternatives. Through a literature search, shown in Appendix~1\ref{app:augmentation}, it was found that only a single paper was published which performed dataset augmentation on structural connectomics data, specifically for the classification of multiple sclerosis patients on binarized connectivity graphs~\cite{barile2021data}. This work adapted a generative adversarial network (GAN) architecture, a structure that is commonly used in other domains of data augmentation. Published applications using GANs or other deep learning approaches for image data augmentation have reported increases in classification accuracy from $0.01-0.1$~\cite{shorten2019survey}. Given the average improvement in performance of $0.04$ presented here, MCA-based augmentation can be considered consistent with the other types of data augmentation when used for classification. The data-agnostic and training-free nature of the MCA approach also enables this technique to be combined with other forms of dataset enhancement, such as directly in GANS or through the addition of realistic noise, when available. MCA has not been tested for other common applications of data augmentation, such as object detection or segmentation.

\section{Conclusion}

This work demonstrates the benefit of augmenting datasets through numerical perturbations. We present an approach which leverages the numerical instability inherent to pipelines for creating more accurate and generalizable classifiers. While the approach and results demonstrated here were specifically relevant in the context of brain imaging, the data-agnostic method for inducing perturbations and off-the-shelf machine learning techniques used suggest that this approach may be widely applicable across domains. This work uniquely shows that numerical uncertainty is an asset which can be harnessed to increase the signal captured from datasets and lead to more robust learned relationships.

\section*{data \& code availability}
The perturbed connectomes were publicly available in a data resource previously produced and made available by the authors~\cite{Kiar2020-yz}. They can be found persistently at \url{https://doi.org/10.5281/zenodo.4041549}, and are made available through The Canadian Open Neuroscience Platform (\url{https://portal.conp.ca/search}, search term ``Kiar''). All software developed for processing or evaluation is publicly available on GitHub at \url{https://github.com/gkpapers/2021AggregateMCA}. Experiments were launched on Compute Canada's HPC cluster environment.

\section*{author contributions}
GK was responsible for the experimental design, data processing, analysis, interpretation, and the majority of writing. All authors contributed to the revision of the manuscript. TG and ACE contributed to experimental design, analysis, interpretation. The authors declare no competing interests for this work. Correspondence and requests for materials should be addressed to Gregory Kiar at \url{gregory.kiar@childmind.org}.

\section*{acknowledgements}
This research was financially supported by the Natural Sciences and Engineering Research Council of Canada (NSERC) (award no. CGSD3-519497-2018). This work was also supported in part by funding provided by Brain Canada, in partnership with Health Canada, for the Canadian Open Neuroscience Platform initiative.

\section*{conflict of interest}
The authors declare no competing interests in the planning, execution, curation, presentation, or publication of this work.

% Submissions are not required to reflect the precise reference formatting of the journal (use of italics, bold etc.), however it is important that all key elements of each reference are included.
\bibliography{main}

\clearpage
\onecolumn
\section*{Appendix 1: Dataset Augmentation in Diffusion MRI Connectomics}
\label{app:augmentation}

To demonstrate the sparsity of existing applications and techniques which have been developed for dataset augmentation in network neuroscience, a PubMed query (Figure~\ref{pubmedquery}) was run and the results were explored. The query was performed on May 28th, 2021, and the results are discussed below with the complete results file published in this project's GitHub repository. The results of this query indicated:

\begin{itemize}
\item \textbf{13} total papers which were flagged as relevant.
\item \textbf{5} of which were related to dataset augmentation, with the other 8 having been incorrectly flagged as relevant and were either focused on clinical practice, non-human data, or analyzed relationships between structural and functional networks.
\item \textbf{4} of which were related to some modality of MRI, the other 1 focusing on high resolution CT imaging.
\item \textbf{1} of which was focused on improving classification performance through dataset augmentation, with the remaining 3 either centered on applications in transfer learning across datasets of different quality, leveraging T1w MRI data for age prediction, or improving segmentation of MR perfusion images of stroke patients.
\end{itemize}

The sole paper~\cite{barile2021data} which closely mirrored the focal application and demonstrates how a novel generative adversarial network was used to improve the classification of multiple sclerosis patients from healthy controls. This paper demonstrated an improvement in F1 score from $0.66$ to $0.81$. This technique used binary structural connectivity networks for disease prediction using a dataset containing $48$ patients.

\begin{figure}[h!]
\begin{tcolorbox}\centering
((diffusion MRI) OR (diffusion weighted imaging) OR (dwi) OR (dMRI) OR (d-MRI) OR (diffusion Imaging))\\
AND (brain)\\
AND ((network) OR (connectome) OR (structural connectome))\\
AND ((data augmentation) OR (dataset augmentation))
\end{tcolorbox}
\caption{Demonstrative PubMed query to identify papers which apply dataset augmentation in network neuroscience.}
\label{pubmedquery}
\end{figure}

\clearpage
\onecolumn
\section*{Appendix 2: Augmentation Improvement in a BMI Classification Task}

To validate that the observed benefit of using Monte Carlo Arithmetic-derived networks wasn't only found in the specific age classification task tested, we reproduced a body mass index (BMI) classification task~\cite{Park2015-uj}. The experimental configuration shown here was identical to the above, with the class labels assigned via a threshold of $BMI > 25$, which is consistent with examples from the literature~\cite{Park2015-uj}.

One key distinction from the age classification task was that the majority of models in the BMI classification task did not perform statistically better than chance using the reference networks. While $36/40$ combinations of network preprocessing, dimensionality reduction, and classifier type produced a model which was meaningfully better than chance for the age classification task without dataset augmentation, only $10/40$ reference models were significant for the BMI task. The performance for all models was still included in the following analysis in the interest of consistency.

\begin{table}[h!]
\centering
\caption{Statistically significant change in performance for body mass index classification. Red values indicate
significant decline in performance, black values indicate improvement, and empty cells indicate no change. A single
star represents $p < 0.05$, and each additional star is an additional order of magnitude of significance.}
\label{tab1:perf_app}
\small
\begin{tabular}{rccc}
\textbf{Dataset Sampling}  & \textbf{Validation} &      \textbf{Test} & \textbf{Generalizability} \\
\hline
\textbf{Truncate}          &                     &    {\color{red} *} &                       *** \\
\textbf{Jackknife}         &                     &  {\color{red} ***} &                       *** \\
\textbf{Mean}              &                     &    {\color{red} *} &                         * \\
\textbf{Median}            &                     &    {\color{red} *} &                        ** \\
\textbf{Consensus}         &                     &  {\color{red} ***} &                        ** \\
\textbf{Mega-Analysis}     &                ***  &                *** &                         * \\
\textbf{Meta-Analysis}     &                ***  &                *** &                        ** \\
\end{tabular}

\end{table}

\begin{figure}[htbp]\centering
\includegraphics[width=0.8\linewidth]{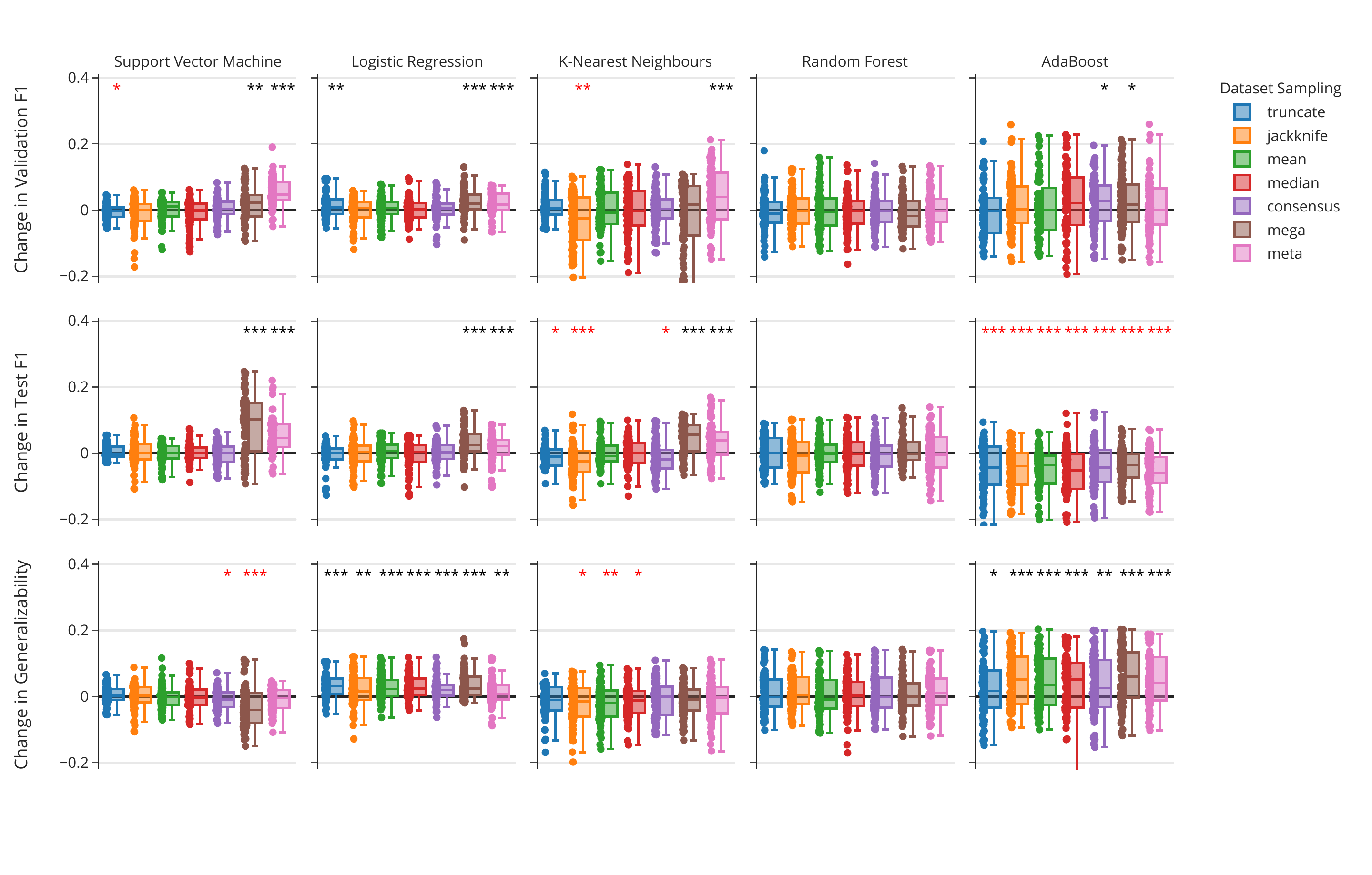}
\caption{Relative change in classifier performance with respect to classifier type and dataset sampling strategies as measured by change in F1 score on the validation set (top) or test set (middle), as well as the generalizability of performance (bottom) for the body mass index classification task. Each star annotation indicates an order of magnitude of statistically significant change, beginning at $0.05$ for one star and decreasing from there, with those in black or red indicating an increase or decrease due to resampling, respectively.}
\label{fig:overall_perf_app}
\end{figure}

\begin{figure}[hbtp]\centering
\includegraphics[width=0.8\linewidth]{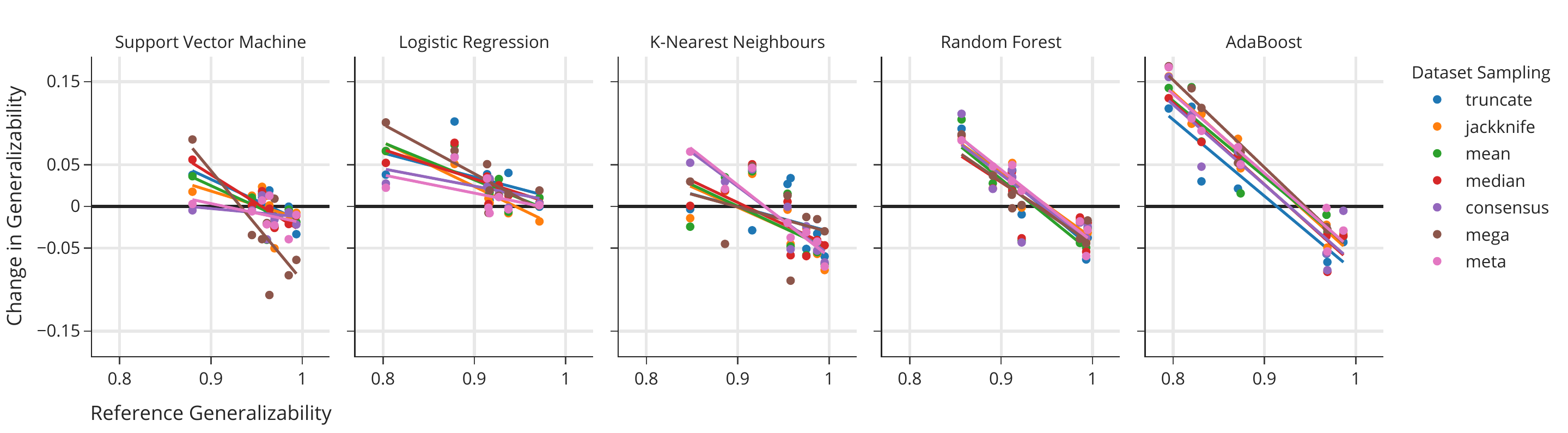}
\includegraphics[width=0.8\linewidth]{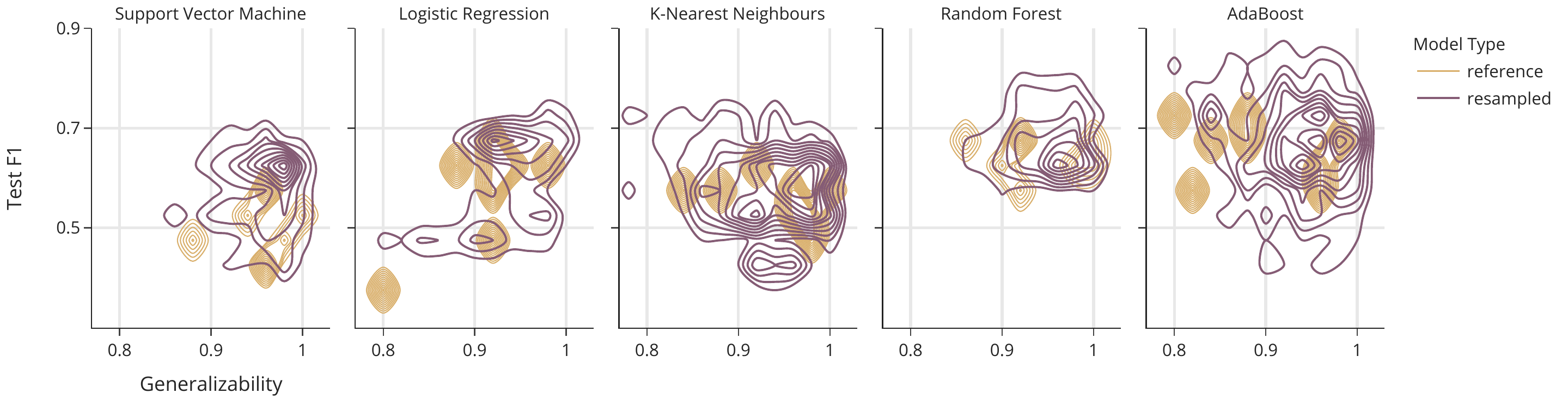}
\caption{Relationship between generalizability and resampling. Top: change in the generalizability of classifiers with respect to the reference generalizability. Each data point represents the mean change in generalizability for all models using the same preprocessing and dimensionality reduction techniques for a given classifier and dataset sampling strategy. Bottom: contour density distributions of generalizability and F1 scores across all models for both reference and resampled training.
% The vertical lines for each plot represent the mean generalizability for the associated models.
}
\label{fig:change_in_gen_app}
\end{figure}

Table~\ref{tab1:perf_app} shows statistically significant change for each dataset augmentation technique aggregated across all models used for body mass index classification. Similarly to the results presented in the main body of this article, the mega- and meta-analytic techniques were superior to the others, and resulted in improvements in model performance both within and out of sample, alongside improving the generalizability in performance. For these two techniques, the performance improved from $0.590$ to $0.623$ on average, with an improvement in generalizability from $0.921$ to $0.939$. In the case of the other dataset sampling strategies, it was noted that the performance decreased while the generalizability increased for each configuration. This suggests that models were ``over-performing'' prior to the data augmentation, and resampling networks served as a regularization. When considering these summaries alongside Figure~\ref{fig:overall_perf_app}, we can see that the noted decrease in test performance was likely driven by the significant reduction in performance of the AdaBoost models while noting an improvement of generalizability suggests a previous over-performance of models built using this classifier type. Figure~\ref{fig:change_in_gen_app} also shows consistent results to those presented above, in which the generalizability trends towards a ``central'' score, and that the dataset sampling approaches tend to shift the model performance either towards increased F1, generalizability, or both.

The classifiers and more generally model combinations which performed the best given dataset augmentation via MCA varied across the two experiments, and it is important to note that this was expected and does not change the significance of the presented findings. The purpose of this work was not to design an ideal experimental configuration that was uniformly out-performing another, but demonstrate the efficacy of MCA as a dataset augmentation technique, in particular for domains in which alternatives may not exist more generally.

\graphicalabstract{figures/0.pdf}{This paper demonstrates how Monte Carlo Arithmetic, a data-agnostic perturbation technique, can be used for dataset augmentation during the generation of structural connectomes and improve downstream phenotypic prediction.}

\end{document}